\documentclass[aps,prl,twocolumn,superscriptaddress]{revtex4-1}
\pdfoutput=1
\usepackage{graphicx}
\usepackage{amsmath,amsfonts,amssymb}
\usepackage[colorlinks=true,linkcolor=blue,citecolor=blue]{hyperref}
\usepackage{siunitx}

\begin{document}
\title{New generation of moir\'e superlattices in doubly aligned hBN/graphene/hBN heterostructures}

\author{Lujun Wang}
\email{lujun.wang@unibas.ch}
\affiliation{Department of Physics, University of Basel, Klingelbergstrasse 82, CH-4056 Basel, Switzerland}

\author{Simon Zihlmann}
\affiliation{Department of Physics, University of Basel, Klingelbergstrasse 82, CH-4056 Basel, Switzerland}

\author{Ming-Hao Liu}
\affiliation{Department of Physics, National Cheng Kung University, Tainan 70101, Taiwan}

\author{P\'eter Makk}
\affiliation{Department of Physics, University of Basel, Klingelbergstrasse 82, CH-4056 Basel, Switzerland}
\affiliation{Department of Physics, Budapest University of Technology and Economics and Nanoelectronics Momentum Research Group of the Hungarian Academy of Sciences, Budafoki ut 8, 1111 Budapest, Hungary}

\author{Kenji Watanabe}
\affiliation{National Institute for Material Science, 1-1 Namiki, Tsukuba, 305-0044, Japan}

\author{Takashi Taniguchi}
\affiliation{National Institute for Material Science, 1-1 Namiki, Tsukuba, 305-0044, Japan}

\author{Andreas Baumgartner}
\email{andreas.baumgartner@unibas.ch}
\affiliation{Department of Physics, University of Basel, Klingelbergstrasse 82, CH-4056 Basel, Switzerland}

\author{Christian Sch\"onenberger}
\affiliation{Department of Physics, University of Basel, Klingelbergstrasse 82, CH-4056 Basel, Switzerland}

\begin{abstract}
The specific rotational alignment of two-dimensional lattices results in a moir\'e superlattice with a larger period than the original lattices and allows one to engineer the electronic band structure of such materials. So far, transport signatures of such superlattices have been reported for graphene/hBN and graphene/graphene systems. Here we report moir\'e superlattices in fully hBN encapsulated graphene with both the top and the bottom hBN aligned to the graphene. In the graphene, two different moir\'e superlattices form with the top and the bottom hBN, respectively. The overlay of the two superlattices can result in a third superlattice with a period larger than the maximum period (\SI{14}{nm}) in the graphene/hBN system, which we explain in a simple model. This new type of band structure engineering allows one to artificially create an even wider spectrum of electronic properties in two-dimensional materials.
\end{abstract}

\maketitle

Superlattice (SL) structures have been used to engineer electronic properties of two-dimensional electron systems for decades \cite{Weiss1989,Weiss1991,Pfannkuche1992,Ferry1992,Schloesser1996,Albrecht1999,Albrecht2001,Geisler2004}. Due to the peculiar electronic properties of graphene \cite{CastroNeto2009}, SLs in graphene are of particular interest \cite{Park2008a,Park2008,Barbier2008,Brey2009,Sun2010,Burset2011,Ortix2012} and have been investigated extensively utilizing different approaches, such as electrostatic gating \cite{Dubey2013,Drienovsky2014,Drienovsky2018}, chemical doping \cite{Sun2011}, etching \cite{Bai2010,Sandner2015,Yagi2015}, lattice deformation \cite{Zhang2018} and surface dielectric patterning \cite{Forsythe2018}. Since the introduction of hexagonal boron nitride (hBN) as a substrate for graphene electronics \cite{Dean2010}, moir\'e superlattices (MSLs) originating from the rotational alignment of the two lattices have been first observed and studied by STM \cite{Decker2011,Xue2011,Yankowitz2012}. It then triggered many theoretical \cite{Kindermann2012,Wallbank2013,Song2013,Moon2014} and experimental studies, where secondary Dirac points \cite{Ponomarenko2013,Dean2013,Hunt2013}, the Hofstadter Butterfly \cite{Ponomarenko2013,Dean2013,Hunt2013,Yu2014,Wang2015a}, Brown-Zak oscillations \cite{Ponomarenko2013,KrishnaKumar2017}, the formation of valley polarized currents \cite{Gorbachev2014} and many other novel electronic device characteristics \cite{Woods2014,Shi2014,Lee2016,Wang2016,Handschin2017,Spanton2018} have been observed. 

Recently, another interesting graphene MSL system has drawn considerable attention -- twisted bilayer graphene, where two monolayer graphene sheets are stacked on top of each other with a controlled twist angle. For small twist angles, insulating states \cite{Cao2016}, strong correlations \cite{Kim2017} and a network of topological channels \cite{Rickhaus2018} have been reported experimentally. More strikingly, unconventional superconductivity \cite{Cao2018a,Yankowitz2018a} and Mott-like insulator states \cite{Cao2018,Yankowitz2018a} have been achieved, when the twist angle is tuned to the so-called ``magic angle", where the electronic band structure near zero Fermi energy becomes flat, due to the strong interlayer coupling.

\begin{figure}[htb]
	\centering
	\includegraphics[width=8.46cm]{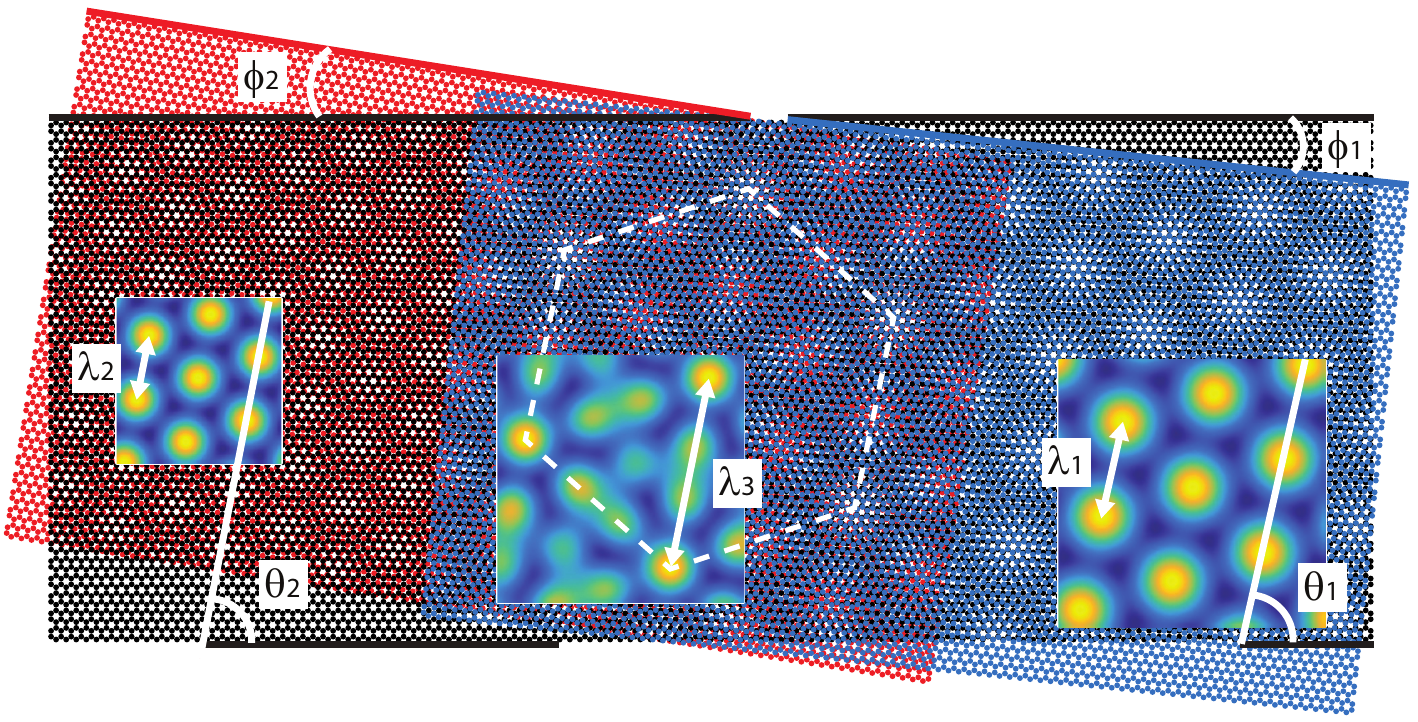}
	\caption{Illustration of three different MSLs formed in a hBN/graphene/hBN heterostructure. Blue, black and red hexagonal lattices represent top hBN, graphene and bottom hBN lattices, respectively. $\phi_1$ ($\phi_2$) is the twist angle between top (bottom) hBN and graphene. $\theta_1$ ($\theta_2$) indicates the orientation of the corresponding MSL with respect to graphene. The resulting moir\'e periods are indicated with $\lambda_{1,2,3}$. The 3L-MSL (middle part) has a larger period than both 2L-MSLs (left and right parts). Insets: moir\'e potential calculations.}
	\label{fig:illustration}
\end{figure}

So far, MSL engineering in graphene has concentrated mostly on MSLs based on two relevant layers (2L-MSLs). However, graphene necessarily forms two interfaces, namely at the top and at the bottom, which can result in a much richer and more flexible tailoring of the graphene band structure. Due to the 1.8\% larger lattice constant of hBN, the largest possible moir\'e period that can be achieved in graphene/hBN systems is limited to about $\SI{14}{\nano\meter}$ \cite{Yankowitz2012}, which occurs when the two layers are fully aligned. This situation changes when both hBN layers are aligned to the graphene layer. Here, we report the observation of a new MSL which can be understood by the overlay of two 2L-MSLs that form between the graphene monolayer and the top and bottom hBN layers of the encapsulation stack, respectively. Figure \ref{fig:illustration} illustrates the formation of the MSLs when both hBN layers are considered. On the right side of the illustration, only the top hBN (blue) and the graphene (black) are present, which form the top 2L-MSL with period $\lambda_1$. The bottom hBN (red) forms the bottom 2L-MSL with graphene, shown on the left with period $\lambda_2$. In the middle of the illustration all three layers are present and a new MSL (3L-MSL) forms with a longer period, indicated with $\lambda_3$. The influence of the MSL can be modeled as an effective periodic potential with the same symmetry. The periodic potentials for the top 2L-MSL and the bottom 2L-MSL are calculated following the model introduced in Ref \cite{Yankowitz2012}, shown as insets in Fig. \ref{fig:illustration}. To calculate the potentials for the 3L-MSL, we sum over the periodic potentials of the top 2L-MSL and the bottom 2L-MSL. The period of the 3L-MSL from the potential calculation matches very well the one of the lattice structure in the illustration. In the transport measurements, we demonstrate that MSL with a period longer than $\SI{14}{nm}$ can indeed be obtained in doubly aligned hBN/graphene/hBN heterostructures, coexisting with the graphene/hBN 2L-MSLs. These experiments are in good agreement with a simple model for the moir\'e periods for doubly aligned hBN/graphene/hBN devices.

\begin{figure}[htb]
	\centering
	\includegraphics[width=8.46cm]{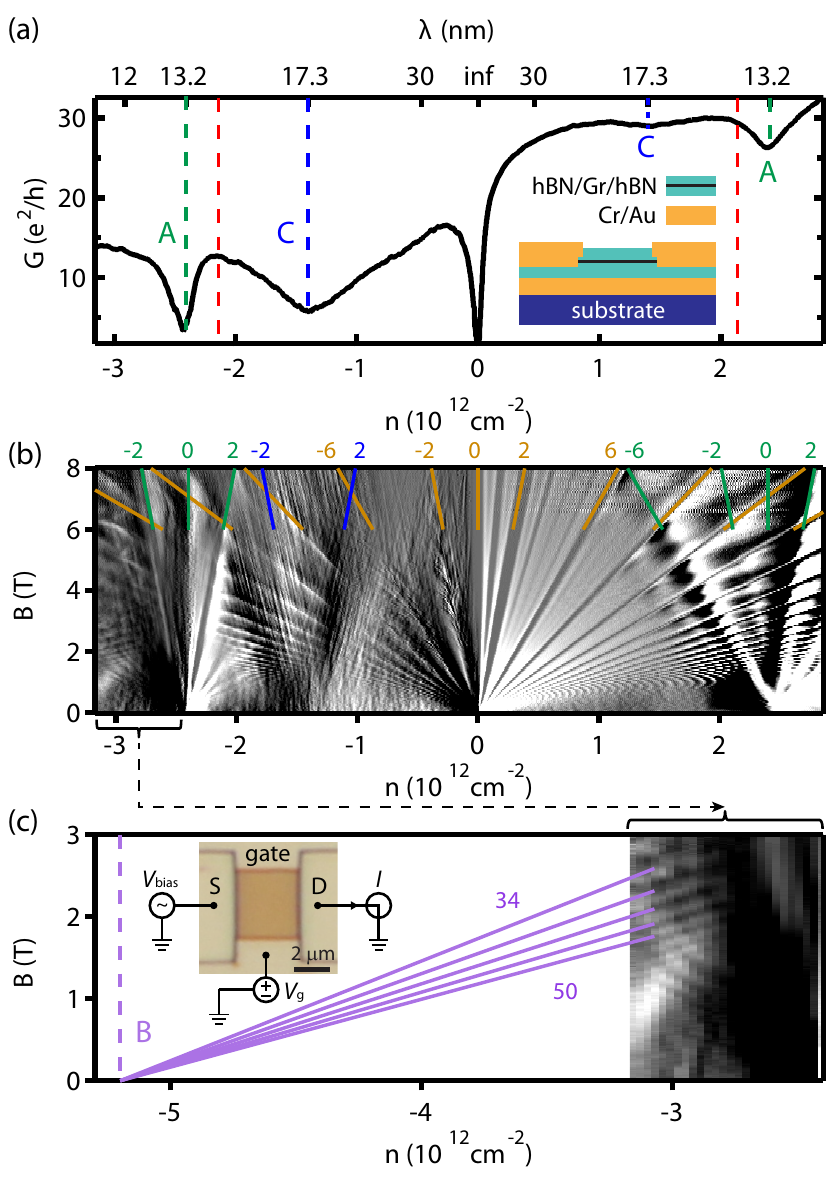}
	\caption{Electronic transport at $\SI{4.2}{\kelvin}$. \textbf{(a)} Two-terminal differential conductance $G$ as a function of charge carrier density $n$. In addition to the MDP, there are 4 other conductance minima at $n_{s_A} \approx \SI{\pm 2.4e12}{\per\square\centi\meter}$ (green dashed lines) and $n_{s_C} \approx \SI{\pm 1.4e12}{\per\square\centi\meter}$ (blue dashed lines), respectively. The top axis shows the moir\'e periods $\lambda = \sqrt{4\pi/3n_s}$. The red dashed lines indicate the longest period (lowest density) for a graphene/hBN MSL. Inset: schematic of the cross section of our device. \textbf{(b)} $\mathrm{d}G/\mathrm{d}n$ as a function of $n$ and $B$ of the same device. Filling factors fan out from all DPs, except for the blue one on the electron side, and are indicated on top of the diagram, calculated as $\nu \equiv nh/(eB)$, where $n$ is counted from each DP. \textbf{(c)} Zoom-in on the left side of \textbf{(b)}. There are additional lines fanning out from an even higher density $n_{s_B} \approx \SI{5.2e12}{\per\square\centi\meter}$, labeled B. The filling factors of these lines are 34, 38, 42, 46 and 50, respectively. Inset: micrograph and experimental setup of the presented device. `S' and `D' are the source and drain contacts, respectively.} 
	\label{fig:fan_diagram}
\end{figure}

We fabricated fully encapsulated graphene devices with both the top and the bottom hBN layers aligned to the graphene using a dry-transfer method \cite{Wang2013}. We estimate an alignment precision of $\SI{\sim 1}{\degree}$. A global metallic bottom gate is used to tune the charge carrier density $n$, and one-dimensional Cr/Au edge contacts are used to contact the graphene \cite{Wang2013} (see inset of Fig. \ref{fig:fan_diagram}(a)). Transport measurements were performed at $\SI{4.2}{\kelvin}$ using standard low-frequency lock-in techniques. 

The two-terminal differential conductance, $G$, of one device, shown as inset of Fig. \ref{fig:fan_diagram}(c), is plotted as a function of $n$ in Fig. \ref{fig:fan_diagram}(a) (data from other devices with similar characteristics, including bilayer graphene devices, are presented in the Supporting Information). The charge carrier density $n$ is calculated from the gate voltage using a parallel plate capacitor model. The average conductance is lower on the hole side ($n < 0$) than on the electron side ($n > 0$), which we attribute to n-type contact doping resulting in a p-n junction near the contacts. The sharp dip in conductance at $n = 0$ is the main Dirac point (MDP) of the pristine graphene. Our device shows a large field-effect mobility of $\SI{\sim 90000}{\square\centi\meter\per\volt\per\second}$, extracted from a linear fit around the MDP. The residual doping is of the order $\delta n \approx \SI{1e10}{\per\square\centi\meter}$, extracted from the width of the MDP. In addition to the MDP, we find two pairs of conductance minima symmetrically around the MDP at higher doping, labeled A and C, which we attribute to two MSLs. The minima on the hole side are more pronounced than their counterparts on the electron side, similar to previously reported MSLs \cite{Yankowitz2012,Ponomarenko2013,Dean2013,Hunt2013}.

Superlattice Dirac points (SDPs) are expected to form at the superlattice Brillouin zone boundaries at $\mathbf{k} = \mathbf{G}/2$, where $\left | \mathbf{G}  \right | =4\pi/(\sqrt{3}\lambda)$ is the length of the superlattice wavevector and $\lambda$ the moir\'e period \cite{Park2008}. For graphene, $k$ is related to $n$ by $k = \sqrt{\pi n}$. The position of the SDPs in charge carrier density for a given period $\lambda$ is then $n_s = 4 \pi/(3 \lambda^2)$. The pair of conductance minima at $n_{s_A} \approx \SI{\pm 2.4e12}{\per\square\centi\meter}$ can be explained by a graphene/hBN 2L-MSL with a period of about $\SI{13.2}{nm}$. However, the pair of conductance minima at $n_{s_C} \approx \SI{\pm 1.4e12}{\per\square\centi\meter}$ cannot be explained by a single graphene/hBN 2L-MSL, since it corresponds to a superlattice period of about $\SI{17.3}{nm}$, clearly larger than the maximum period of $\SI{\sim 14}{nm}$ in a graphene/hBN moir\'e system. We attribute the presence of the conductance dips at $n_{s_C}$ to a new MSL that is formed by the three layers together: top hBN, graphene and bottom hBN. This 3L-MSL can have a period considerably larger than $\SI{14}{nm}$.

To substantiate this claim, we now analyse the data obtained in the quantum Hall regime. Figure \ref{fig:fan_diagram}(b) shows the Landau fan of the same device, where the numerical derivative of the conductance with respect to $n$ is plotted as a function of $n$ and the out-of-plane magnetic field $B$. Near the MDP, we observe the standard quantum Hall effect for graphene with plateaus at filling factors $\nu \equiv nh/(eB) = \pm 2, \pm 6, \pm 10, ...$, with $h$ the Planck constant and $e$ the electron charge. This spectrum shows the basic Dirac nature of the charge carriers in graphene. The broken symmetry states occur for $B \geqslant \SI{2}{\tesla}$, suggesting a high device quality. Around the SDPs at $n_{s_A} \approx \SI{\pm 2.4e12}{\per\square\centi\meter}$, the plot also shows filling factors $\nu \equiv (n-n_{s_A})h/(eB) = \pm 2, \pm 6, ...$, consistent with previous graphene/hBN MSL studies \cite{Ponomarenko2013}. Around the SDPs at $n_{s_C} \approx \SI{\pm 1.4e12}{\per\square\centi\meter}$, there are also clear filling factors fanning out on the hole side with $\nu \equiv (n-n_{s_C})h/(eB) = \pm 2$, which is consistent with a Dirac spectrum at $n_{s_C}$, while on the electron side the corresponding features are too weak to be observed. In addition, lines faning out from a SDP located  at density $n < \SI{-3e12}{\per\square\centi\meter}$ are observed. A zoom-in is plotted in Fig. \ref{fig:fan_diagram}(c). The lines extrapolate to a density of about $\SI{-5.2e12}{\per\square\centi\meter}$, denoted $n_{s_B}$, with filling factors $\nu = 34, 38, 42, 46, ...$ This density cannot be explained by the ``tertiary" Dirac point occuring at the density of about $1.65 n_{s_A}$, which comes from a Kekul\'e superstructure on top of the graphene/hBN MSL \cite{Chen2017}. However, $n_{s_B}$ matches the SDP from a MSL with a period of about $\SI{9}{\nano\meter}$. We therefore attribute it to a 2L-MSL originating from the alignment of the second hBN layer to the graphene layer.

\begin{figure}[htb]
	\centering
	\includegraphics[width=8.46cm]{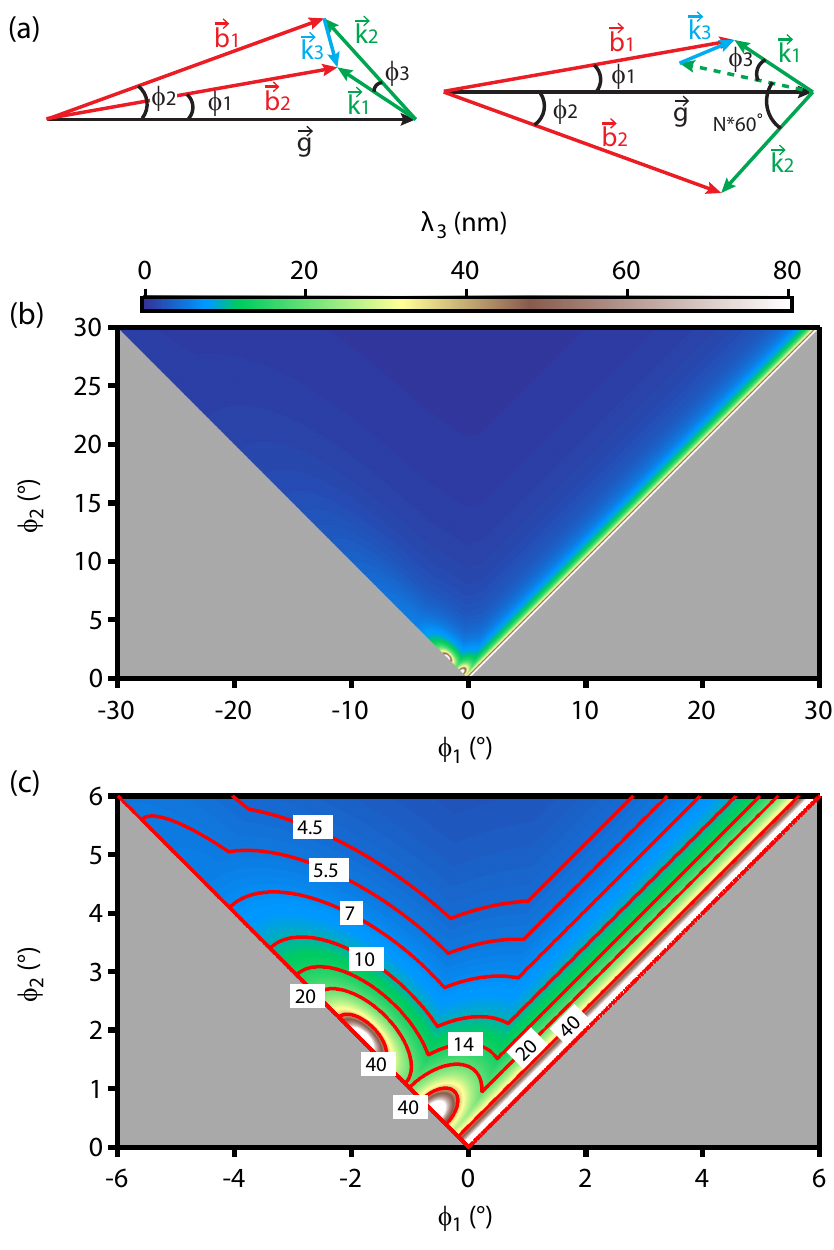}
	\caption{\textbf{(a)} Schematics in reciprocal space for the formation of different MSLs, where $\vec{g}, \vec{b}_1, \vec{b}_2, \vec{k}_1, \vec{k}_2$ and $\vec{k}_3$ are one of the reciprocal lattice vectors for graphene, top hBN, bottom hBN, top 2L-MSL, bottom 2L-MSL and 3L-MSL, respectively. $N$ is an integer, which can be 1, 2, or 3. \textbf{(b)} $\lambda_3$ plotted as a function of $\phi_1$ and $\phi_2$ for all possible twist angles. \textbf{(c)} Zoom-in of \textbf{(b)} for small twist angles. Numbers on the contour lines indicate the values of $\lambda_3$ in nm.}
	\label{fig:lambda3}
\end{figure}

As derived in Ref \cite{Yankowitz2012,Moon2014}, the period $\lambda$ for a graphene/hBN MSL is given by

\begin{equation}\label{eqn:lambda}
\lambda = \frac{(1+\delta)a}{\sqrt{2(1+\delta)(1-\cos{\phi})+\delta^2}},
\end{equation}
\\
where $a$ ($\SI{2.46}{\angstrom}$) is the graphene lattice constant, $\delta$ (1.8\%) is the lattice mismatch between hBN and graphene and $\phi$ (defined for $\SI{-30}{\degree}$ to $\SI{30}{\degree}$) is the twist angle of hBN with respect to graphene. The moir\'e period is maximum at $\phi = 0$ with a value of $\lambda \approx \SI{14}{nm}$. This corresponds to the lowest carrier density of $n_{min} \approx \SI{\pm 2.2e12}{\per\square\centi\meter}$ for the position of the SDPs (red dashed lines in Fig. \ref{fig:fan_diagram}(a)). The orientation of the MSL is described by the angle $\theta$ relative to the graphene lattice,

\begin{equation}\label{eqn:theta}
\tan{\theta} = \frac{-\sin{\phi}}{(1+\delta)-\cos{\phi}}.
\end{equation}
\\
For the graphene/hBN system, one finds $\left | \theta \right | \lesssim \SI{80}{\degree}$ \cite{Yankowitz2012}. These two equations describe the top 2L-MSL and the bottom 2L-MSL, as shown schematically in Fig. \ref{fig:illustration}.

In a fully encapsulated graphene device, not only one, but both hBN layers can be aligned to the graphene layer so that two graphene/hBN 2L-MSLs can form. In this case, the potential modulations of the two 2L-MSLs are superimposed and form a MSL with a third periodicity. The values of the resulting periods can be understood based on Fig. \ref{fig:lambda3}(a). The vectors $\vec{g}$, $\vec{b}_1$ and $\vec{b}_2$ denote one of the reciprocal lattice vectors for the graphene, the top hBN and the bottom hBN layers, respectively. The twist angle between the top (bottom) hBN and graphene is denoted $\phi_1$ ($\phi_2$). Following the derivations in Ref \cite{Yankowitz2012,Moon2014}, one of the top 2L-MSL (bottom 2L-MSL) reciprocal lattice vectors $\vec{k}_1$ ($\vec{k}_2$) is given by the vector connecting $\vec{g}$ to $\vec{b}_1$ ($\vec{b}_2$). The moir\'e period $\lambda_{1,2}$ is then given by $\lambda_{1,2} = 4\pi/\sqrt{3}\left | \vec{k}_{1,2} \right |$, which is explicitly described by Eq. (\ref{eqn:lambda}) as a function of the twist angle $\phi_{1,2}$. Since the reciprocal lattices of the top 2L-MSL and the bottom 2L-MSL are triangular, the same as those for graphene and hBN, we can use the same approach to derive the 3L-MSL, which is described by the vector connecting $\vec{k_2}$ to $\vec{k_1}$, denoted $\vec{k_3}$. The 3L-MSL period is then given by $\lambda_3 = 4\pi/\sqrt{3}\left | \vec{k}_3 \right |$. 

In order to calculate $\lambda_3$ using Eq. (\ref{eqn:lambda}), we first need to find the new $a$, $\delta$ and $\phi$. Due to symmetry, we only consider $\phi_1 < \phi_2$, so $\lambda_2$, the smaller period of the two graphene/hBN 2L-MSLs, becomes the new $a$ and the new $\delta$ will then be given by $(\lambda_1 - \lambda_2) / \lambda_2$. The new $\phi$, denoted $\phi_3$, is determined by $\left | \theta_1 - \theta_2\right |$, where $\theta_1$ ($\theta_2$) is the relative orientation of the top 2L-MSL (bottom 2L-MSL) with respect to the graphene lattice, described by Eq. (\ref{eqn:theta}). Different cases occur for $\phi_3$ due to the $\SI{60}{\degree}$ rotational symmetry of the lattices. Since $\phi$ in Eq. (\ref{eqn:lambda}) is defined for $\SI{-30}{\degree}$ to $\SI{30}{\degree}$, we subtract multiples of $\SI{60}{\degree}$ to bring $\phi_3$ to this range if it is larger than $\SI{30}{\degree}$, given as

$$\phi_3 = \begin{cases}
 \left | \theta_1 - \theta_2\right |& \text{ if } 0 < \left | \theta_1 - \theta_2\right | \leqslant  \SI{30}{\degree}\\ 
 \left | \theta_1 - \theta_2\right | - \SI{60}{\degree} & \text{ if }  \SI{30}{\degree} < \left | \theta_1 - \theta_2\right | \leqslant  \SI{90}{\degree} \\ 
 \left | \theta_1 - \theta_2\right | - \SI{120}{\degree} & \text{ if } \SI{90}{\degree} < \left | \theta_1 - \theta_2\right | \leqslant  \SI{150}{\degree} \\ 
 \left | \theta_1 - \theta_2\right | - \SI{180}{\degree} & \text{ if } \SI{150}{\degree} < \left | \theta_1 - \theta_2\right | \leqslant  \SI{180}{\degree}.
\end{cases}$$
\\
For the first case, the 3L-MSL is effectively the MSL formed by the two hBN layers, as illustrated in the left panel of Fig. \ref{fig:lambda3}(a). Another case is shown in the right panel, where multiples of $\SI{60}{\degree}$ are subtracted, which is equivalent to choosing another reciprocal lattice vector for $\vec{k}_2$ so that it makes an angle within $\SI{\pm30}{\degree}$ with $\vec{k}_1$. 

Figure \ref{fig:lambda3}(b) plots all possible values for $\lambda_3$, as a function of $\phi_1$ and $\phi_2$, by using Eq. (\ref{eqn:lambda}) with the new parameters. Theoretically $\lambda_3$ varies from below $\SI{1}{\nano\meter}$ to infinity, but one finds values larger than $\SI{14}{\nano\meter}$ only for small twist angles (see Fig. \ref{fig:lambda3}(c)). For most angles $\lambda_3$ is very small, which explains why MSLs with periods larger than $\SI{14}{nm}$ have not been reported in previous studies, where only one hBN layer was aligned intentionally to the graphene layer. 

Most of Fig. \ref{fig:lambda3}(c) can be understood intuitively. On the line of the right diagonal with $\phi_1 \equiv \phi_2$, we have $\lambda_1 = \lambda_2$ and $\theta_1 = \theta_2$, therefore $\phi_3 = 0$, which results in $\lambda_3 = \infty$. This case is similar to the twisted bilayer graphene with a twist angle of 0, which does not form a MSL (or a MSL with infinitely large period). On the diagonal line in the left part with $\phi_1 \equiv -\phi_2$, one has $\lambda_1 = \lambda_2$, but $\theta_1 = -\theta_2$. Therefore $\phi_3$ can have non-zero values, resulting in different $\lambda_3$ values. This case is again similar to the twisted bilayer graphene, but with a tunable twist angle. The two maxima in $\lambda_3$ occur if $\left | \theta_1 - \theta_2\right | = \SI{60}{\degree}$, $\SI{120}{\degree}$, where $\phi_3$ is reset to 0 due to symmetry of the lattices, which is equivalent to the diagonal on the right part. The kinks on the contour lines come from the $\SI{60}{\degree}$ rotational symmetry of the lattices, where $\left | \phi_3 \right | = \SI{30}{\degree}$.  

We now compare this simple model to our experiments. From the SDPs at $n_{s_A} \approx \SI{\pm2.4e12}{\per\square\centi\meter}$, we calculate the corresponding moir\'e period $\lambda_1 \approx \SI{13.2}{\nano\meter}$ and the twist angle $\left | \phi_1 \right | \approx \SI{0.34}{\degree}$. Similarly, for the extrapolated SDP at $n_{s_B} \approx \SI{-5.2e12}{\per\square\centi\meter}$, we obtain $\lambda_2 \approx \SI{9}{\nano\meter}$ and $\left | \phi_2 \right | \approx \SI{1.2}{\degree}$. The two twist angles give us two points in the map in Fig. \ref{fig:lambda3}(c): $\SI{\sim 17.2}{\nano\meter}$ for ($\SI{0.34}{\degree}$, $\SI{1.2}{\degree}$) and $\SI{\sim 27.1}{\nano\meter}$ for ($\SI{-0.34}{\degree}$, $\SI{1.2}{\degree}$). The $\SI{\sim 17.2}{\nano\meter}$ matches very well the value $\SI{\sim 17.3}{\nano\meter}$ extracted from the new-generation SDPs at $n_{s_C} \approx \SI{\pm1.4e12}{\per\square\centi\meter}$ in the transport measurement, which confirms that the new-generation SDPs come from the 3L-MSL. 

We fabricated five hBN/graphene/hBN heterostructures in total, two of which exhibit 3L-MSL features. Data from devices of the second heterostructure are presented in the Supporting Information, which has a 3L-MSL with $\lambda_3 \approx \SI{29.6}{\nano\meter}$.

In conclusion, we have demonstrated the emergence of a new generation of MSLs in fully encapsulated graphene devices with aligned top and bottom hBN layers. In these devices we find three different superlattice periods, one of which is larger than the maximum graphene/hBN moir\'e period, which we attribute to the combined top and bottom hBN potential modulation. Whereas our model describes qualitatively the densities where these 3L-MSL features occur, the precise nature of the band structure distortions is unknown. The alignment of both hBN layers to graphene opens new possibilities for graphene band structure engineering, therefore providing motivation for further studies. Our new approach of MSL engineering is not limited to graphene with hBN, but applies to two-dimensional materials in general, such as twisted trilayer graphene, graphene with transition metal dichalcogenides, etc., which might open a new direction in ``twistronics" \cite{Carr2017,Ribeiro-Palau2018}.

\section*{Acknowledgement}
This work has received funding from the Swiss Nanoscience Institute (SNI), the ERC project TopSupra (787414), the European Union Horizon 2020 research and innovation programme under grant agreement No. 696656 (Graphene Flagship), the Swiss National Science Foundation, the Swiss NCCR QSIT, Topograph, ISpinText FlagERA network and from the OTKA FK-123894 grants. P.M. acknowledges support from the Bolyai Fellowship, the Marie Curie grant and the National Research, Development and Innovation Fund of Hungary within the Quantum Technology National Excellence Program (Project Nr. 2017-1.2.1-NKP-2017-00001). M.-H.L. acknowledges support from Taiwan Minister of Science and Technology (MOST) under Grant No. 107-2112-M-006-004-MY3. Growth of hexagonal boron nitride crystals was supported by the Elemental Strategy Initiative conducted by the MEXT, Japan and the CREST (JPMJCR15F3), JST. The authors thank David Indolese and Peter Rickhaus for fruitful discussions.

\bibliography{DoubleMoire}

%
%
%

\end{document}


\beginsupplement

\title{Supporting information for \\ New generation of moir\'e superlattices in doubly aligned hBN/graphene/hBN heterostructures}


\author{Lujun Wang}
\email{lujun.wang@unibas.ch}
\affiliation{Department of Physics, University of Basel, Klingelbergstrasse 82, CH-4056 Basel, Switzerland}

\author{Simon Zihlmann}
\affiliation{Department of Physics, University of Basel, Klingelbergstrasse 82, CH-4056 Basel, Switzerland}

\author{Ming-Hao Liu}
\affiliation{Department of Physics, National Cheng Kung University, Tainan 70101, Taiwan}

\author{P\'eter Makk}
\affiliation{Department of Physics, University of Basel, Klingelbergstrasse 82, CH-4056 Basel, Switzerland}
\affiliation{Department of Physics, Budapest University of Technology and Economics and Nanoelectronics Momentum Research Group of the Hungarian Academy of Sciences, Budafoki ut 8, 1111 Budapest, Hungary}

\author{Kenji Watanabe}
\affiliation{National Institute for Material Science, 1-1 Namiki, Tsukuba, 305-0044, Japan}

\author{Takashi Taniguchi}
\affiliation{National Institute for Material Science, 1-1 Namiki, Tsukuba, 305-0044, Japan}

\author{Andreas Baumgartner}
\email{andreas.baumgartner@unibas.ch}
\affiliation{Department of Physics, University of Basel, Klingelbergstrasse 82, CH-4056 Basel, Switzerland}

\author{Christian Sch\"onenberger}
\affiliation{Department of Physics, University of Basel, Klingelbergstrasse 82, CH-4056 Basel, Switzerland}

\maketitle


\section{Fabrication}\label{sec:fab}

We align hBN to graphene by aligning the straight edges of each layer. The probability for each alignment is 50\%, because the straight edges can be along either zigzag or armchair direction. Since we need to align both hBN layers to graphene, the probability drops to 25\%. We think that this is why only two out of five samples exhibit 3L-MSL features in the experiment. The hBN/G/hBN stack is directly placed on the metallic gate, so the bottom hBN acts as the dielectric layer which is usually about 20-40nm thick in our case, resulting in a high gating efficiency.   

\section{Sample a}\label{sec:smaple_a}

\begin{figure}[htb]
    \centering
      \includegraphics[width=\columnwidth]{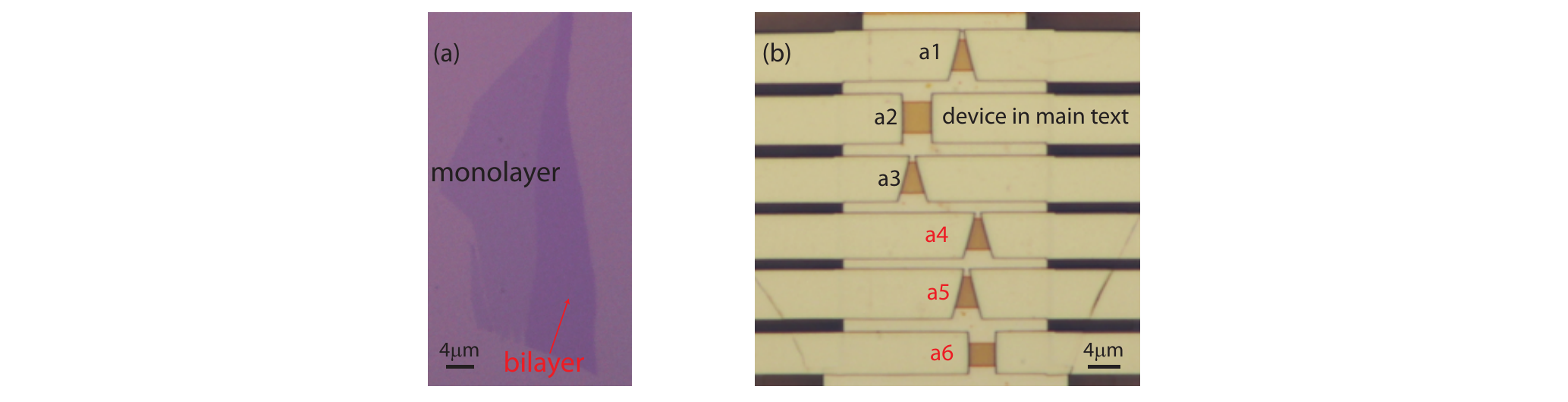}
    \caption{\textbf{(a)} Micrograph of the graphene flake used for sample a. \textbf{(b)} Micrograph of all six finished fully encapsulated devices. a1-a3 are monolayer and a4-a6 are bilayer.}
    \label{fig:image1}
\end{figure}

A flake with both monolayer and bilayer graphene as shown in Fig. \ref{fig:image1}(a) was chosen for sample a. We fabricated six fully encapsulated devices out of this flake, with three monolayer devices and three bilayer devices. The different device geometries are designed for other experiments. The device discussed in the main text is device a2 in Fig. \ref{fig:image1}(b). In the following we show the gate traces of all devices and the Landau fan diagram of monolayer device a3. Unfortunately, we do not have the complete data set for other devices due to a gate leak that appeared during the measurements.

\begin{figure}[htb]
    \centering
      \includegraphics[width=0.8\columnwidth]{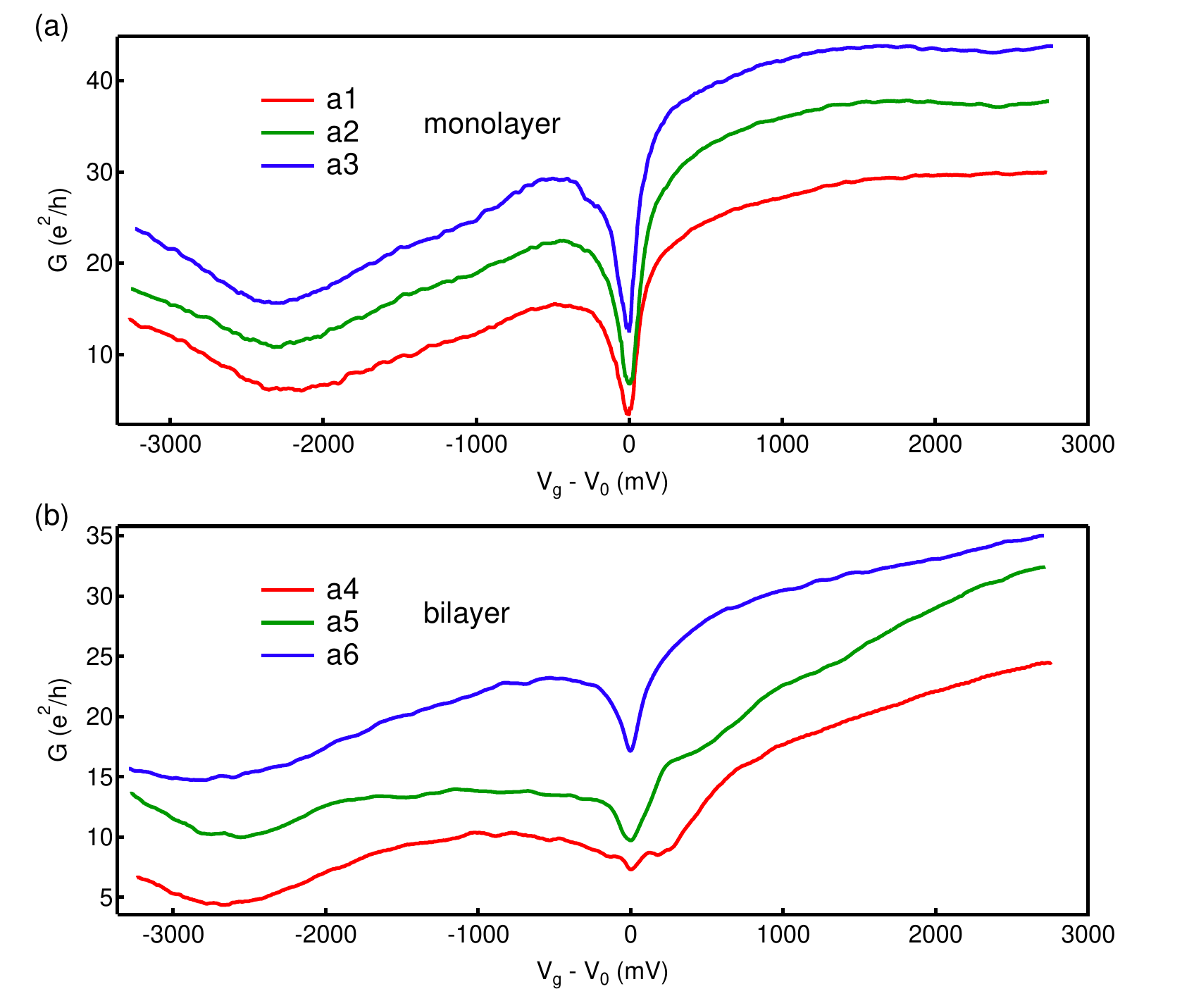}
    \caption{Two-terminal differential conductance $G$ as a function of gate voltage $V_g$ measured at $\SI{4.2}{\kelvin}$ for all three monolayer devices \textbf{(a)} and all three bilayer devices \textbf{(b)}. $V_0$ is around $\SI{250}{\milli\volt}$ for all devices. Curves are shifted by 5$e^2/h$ sequentially in y direction for clarity.}
    \label{fig:CuBJ8_all}
\end{figure}

In Fig. \ref{fig:CuBJ8_all}, two-terminal differential conductance is plotted as a function of gate voltage for all devices. Each curve is offset by an individual $V_0$ in gate voltage in order to shift the MDP to zero gate voltage. All six devices spread over $\SI{50}{\micro\meter}$ show extra conductance minima in addition to the MDP at roughly the same gate voltage, suggesting an intrinsic lattice related origin of these features.

\subsection{Device a3}
\begin{figure}[htb]
    \centering
      \includegraphics[width=0.8\columnwidth]{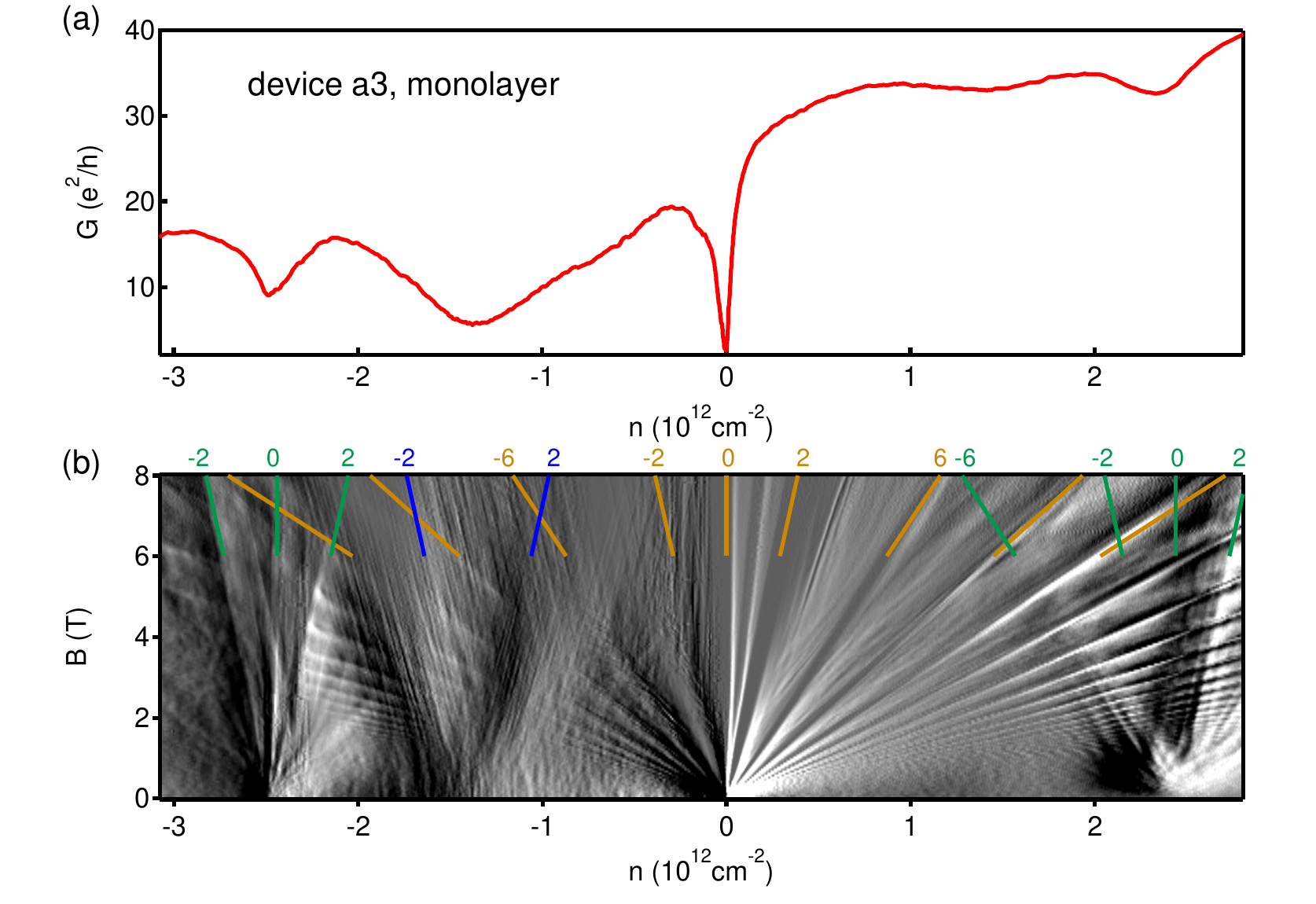}
    \caption{Electronic transport of device a3 at $\SI{4.2}{\kelvin}$. \textbf{(a)} Two-terminal differential conductance $G$ as a function of charge carrier density $n$. \textbf{(b)} $\mathrm{d}G/\mathrm{d}n$ as a function of $n$ and $B$ of the same device. Filling factors are indicated on top of the diagram}
    \label{fig:a3}
\end{figure}

The two-terminal differential conductance of monolayer device a3 is plotted as a function of charge carrier density $n$ in Fig. \ref{fig:a3}(a). In addition to the MDP, two pairs of conductance minima occur symmetrically at $n \approx \SI{\pm 2.4e12}{\per\square\centi\meter}$ and $n \approx \SI{\pm 1.4e12}{\per\square\centi\meter}$, respectively, exactly the same as in device a2 in the main text. The Landau fan diagram (see Fig. \ref{fig:a3}(b)) also looks very similar as that of device a2.

\section{Sample b}\label{sec:sample_b}

\begin{figure}[H]
    \centering
      \includegraphics[width=\columnwidth]{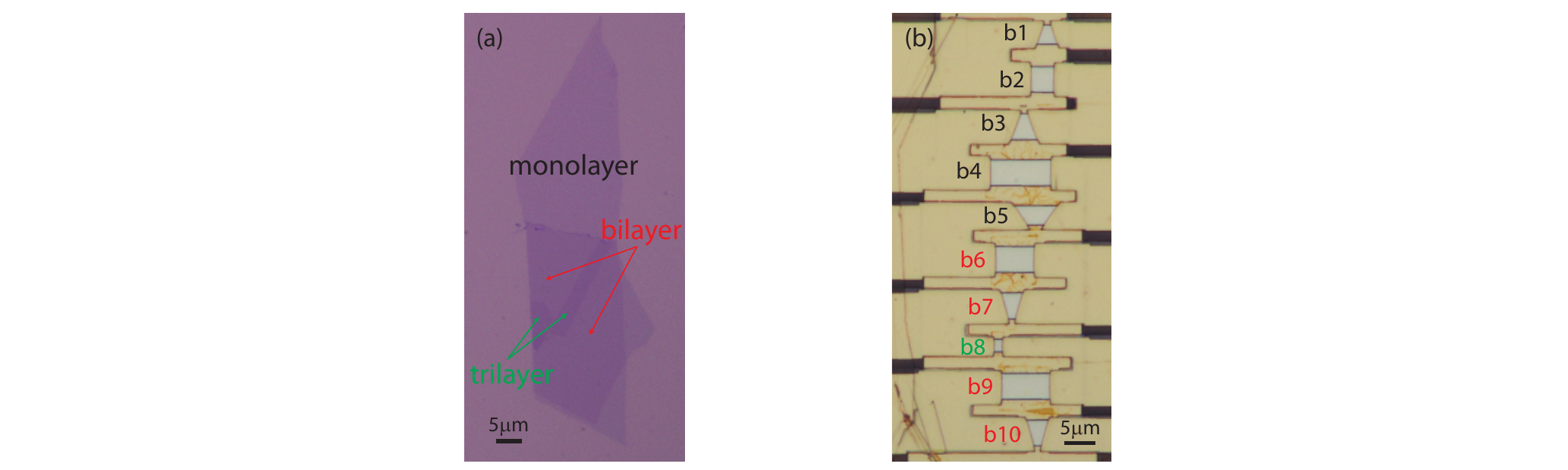}
    \caption{\textbf{(a)} Micrograph of the graphene flake used for sample b. \textbf{(b)} Micrograph of all ten finished fully encapsulated devices. b1-b5 are monolayer, b6, b7, b9, b10 are bilayer and b8 is trilayer.}
    \label{fig:image2}
\end{figure}

\begin{figure}[H]
    \centering
      \includegraphics[width=0.8\columnwidth]{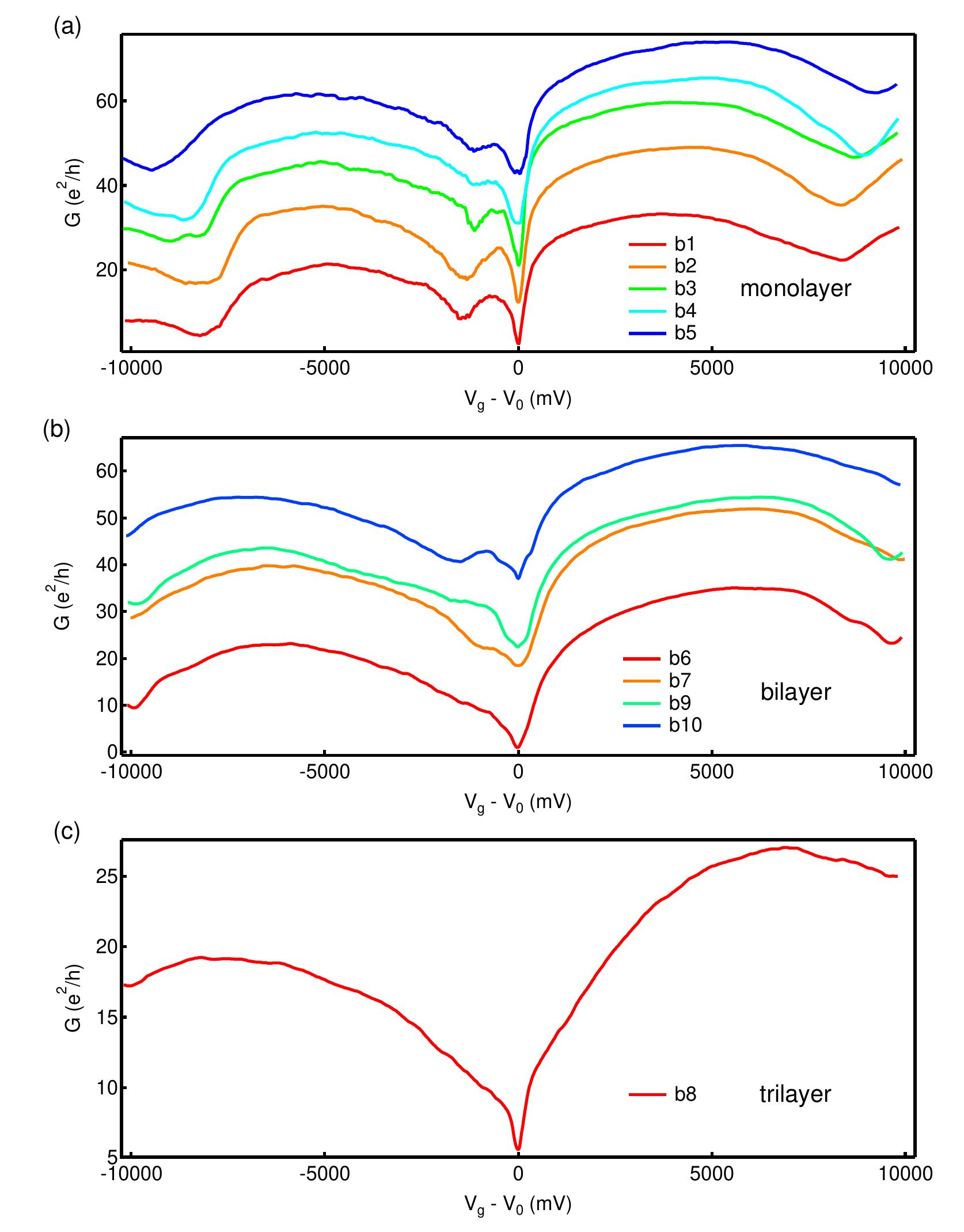}
    \caption{Two-terminal differential conductance $G$ as a function of gate voltage $V_g$ measured at $\SI{4.2}{\kelvin}$ for all five monolayer devices \textbf{(a)}, all four bilayer devices \textbf{(b)} and one trilayer device \textbf{(c)}. $V_0$ varies from $\SI{20}{\milli\volt}$ to $\SI{220}{\milli\volt}$ for different devices. Curves are shifted by 10$e^2/h$ sequentially in y direction for clarity.}
    \label{fig:CuBJ10_all}
\end{figure}

The graphene flake used for sample b is shown in Fig. \ref{fig:image2}(a). Ten fully encapsulated devices were fabricated out of this flake as depicted in Fig. \ref{fig:image2}(b), with five monolayer devices, four bilayer devices and one trilayer device. We show the gate traces of all devices in Fig. \ref{fig:CuBJ10_all} and the Landau fan diagram of one monolayer device b2 in Fig. \ref{fig:b2}. Unfortunately, the gate started to leak during the measurements as one can see for example in Fig. \ref{fig:b2}(b).

Two-terminal differential conductance of all ten devices are plotted as a function of gate voltage in Fig. \ref{fig:CuBJ10_all}. The additional DPs occur at slightly different gate voltages for different devices. One reason for that might be the gate leak, resulting in different lever arms for different devices. The measuring sequence is the same as the labelling, with device b1 measured first and device b10 measured last. The second reason might be a tiny relative rotation of any of the three layers at different locations due to bubbles or ripples formed in the stack during fabrication \cite{Hunt2013}, which leads to effectively slightly different MSLs for different devices. In Fig. \ref{fig:CuBJ10_all}(c), it seems the 3L-MSL DP is almost absent in the trilayer device, which is expected due to the further separation of the top 2L-MSL and the bottom 2L-MSL.
 
\subsection{Device b2}
\begin{figure}[H]
    \centering
      \includegraphics[width=0.8\columnwidth]{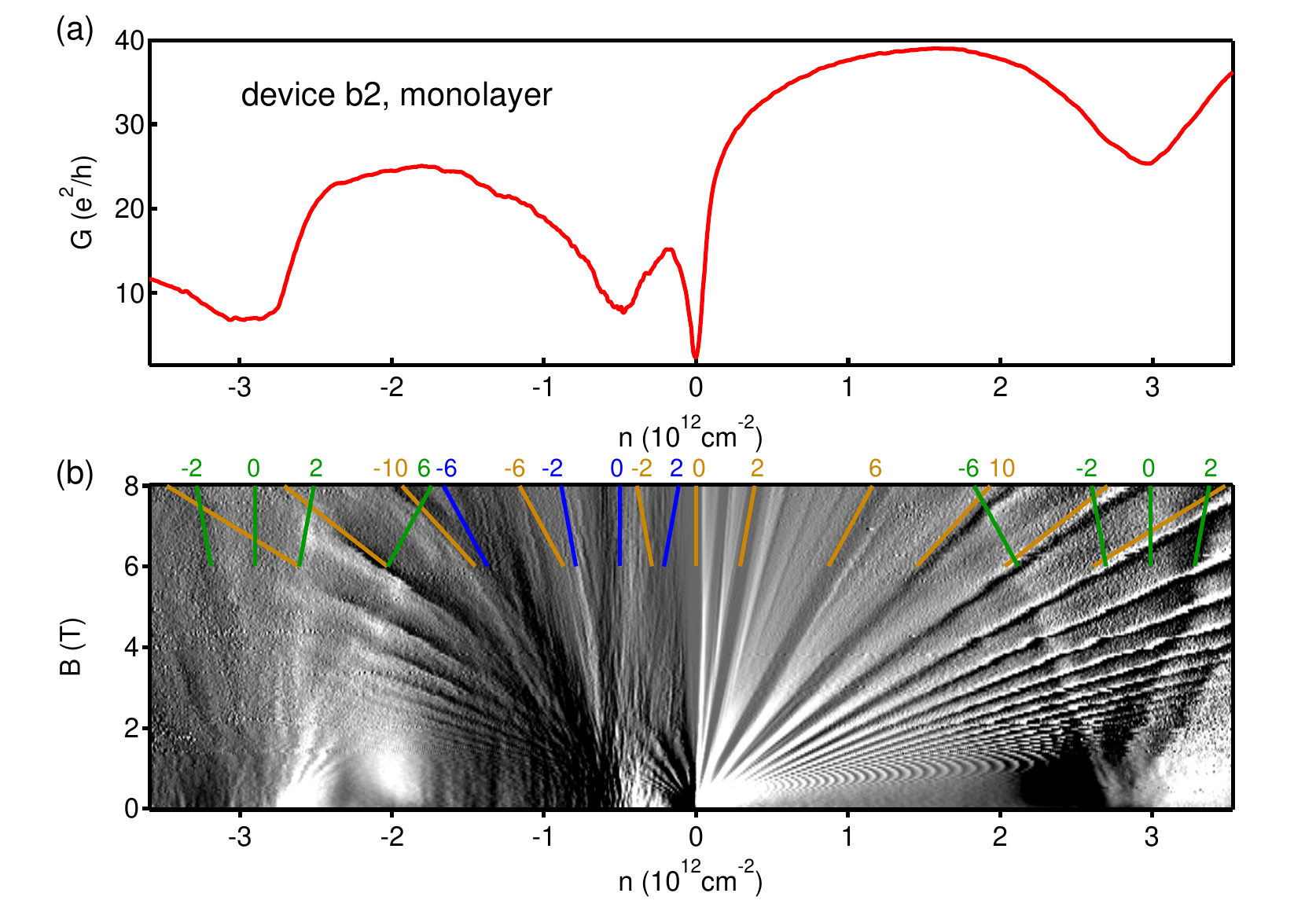}
    \caption{Electronic transport of device b2 at $\SI{4.2}{\kelvin}$. \textbf{(a)} Two-terminal differential conductance $G$ as a function of $n$. \textbf{(b)} $\mathrm{d}G/\mathrm{d}n$ as a function of $n$ and $B$ of the same device. Filling factors are indicated on top of the diagram. The bending and instability/noise of the filling factors at higher densities is due to the gate leak.}
    \label{fig:b2}
\end{figure}

The two-terminal differential conductance of monolayer device b2 is plotted as a function of $n$ in Fig. \ref{fig:b2}(a). In addition to the MDP, one pair of SDPs appear symmetrically at $n \approx \SI{\pm 2.9e12}{\per\square\centi\meter}$, resulting from a graphene/hBN MSL with $\lambda \approx \SI{12}{\nano\meter}$ and $\phi \approx \SI{0.6}{\degree}$. Another SDP appears at $n \approx \SI{-0.48e12}{\per\square\centi\meter}$. There are also filling factors fanning out from this SDP as shown in Fig. \ref{fig:b2}(b). This density corresponds to a superlattice with $\lambda \approx \SI{29.6}{\nano\meter}$, which we attribute to the 3L-MSL. With these parameters and the Fig. 3(c) of the main text, we can deduce back the parameters of the other graphene/hBN MSL to be $\phi \approx \SI{1.1}{\degree}$ or $\SI{-1.1}{\degree}$ and $\lambda \approx \SI{9.5}{\nano\meter}$, corresponding to a density of $n \approx \SI{\pm4.6e12}{\per\square\centi\meter}$. This twist angle is in good agreement with our alignment precision. 

\bibliography{DoubleMoire}